\documentclass[aps,prd,onecolumn,groupedaddress,showpacs,nofootinbib,amssymb
]{revtex4}
\usepackage[dvips]{graphicx}
\usepackage{amssymb}
\usepackage{amsmath}
\usepackage{graphicx}
\usepackage{amsfonts}
\usepackage{bm}
\usepackage{color}

\providecommand{\U}[1]{\protect\rule{.1in}{.1in}}
\newcommand{\be}{\begin{equation}}
\newcommand{\ee}{\end{equation}}

\newcommand{\mincir}{\raise
-3.truept\hbox{\rlap{\hbox{$\sim$}}\raise4.truept\hbox{$<$}\ }}
\newcommand{\magcir}{\raise
-3.truept\hbox{\rlap{\hbox{$\sim$}}\raise4.truept\hbox{$>$}\ }}

\begin{document}

\title{Viable Inflationary Evolution from Einstein Frame Loop Quantum Cosmology}
\author{Jaume de Haro,$^{1}$\,S.D. Odintsov,$^{2,3,4,5}$\,\thanks{odintsov@ieec.uab.es}
V.K. Oikonomou$^{6,7}$\,\thanks{v.k.oikonomou1979@gmail.com}}
\affiliation{
$^{1)}$ Departament de Matem\`atiques, Universitat Polit\`ecnica de Catalunya, Colom 11, 08222 Terrassa, Spain\\
$^{2)}$ ICREA, Passeig Luis Companys, 23, 08010 Barcelona, Spain\\
$^{3)}$ Institute of Space Sciences (ICE,CSIC) C. Can Magrans s/n,
08193 Barcelona, Spain\\
$^{4)}$ Institute of Space Sciences of Catalonia (IEEC),
Barcelona, Spain\\
$^{5)}$ Laboratory for Theoretical Cosmology, Tomsk State University
of Control Systems
and Radioelectronics, 634050 Tomsk, Russia (TUSUR)\\
$^{6)}$Department of Physics, Aristotle University of Thessaloniki, Thessaloniki 54124, Greece\\
$^{7)}$ Tomsk State Pedagogical University, 634061 Tomsk, Russia
}
\tolerance=5000

\begin{abstract}
In this work we construct a bottom-up reconstruction technique for Loop Quantum Cosmology scalar-tensor theories, from the observational indices. Particularly, the reconstruction technique is based on fixing the functional form of the scalar-to-tensor ratio as a function of the $e$-foldings number. The aim of the technique is to realize viable inflationary scenarios, and the only assumption that must hold true in order for the reconstruction technique to work is that the dynamical evolution of the scalar field obeys the slow-roll conditions. We shall use two functional forms for the scalar-to-tensor ratio, one of which corresponds to a popular inflationary class of models, the $\alpha$-attractors. For the latter, we shall calculate the leading order behavior of the spectral index and we shall demonstrate that the resulting inflationary theory is viable and compatible with the latest Planck and BICEP2/Keck-Array data. In addition, we shall find the classical limit of the theory, and as we demonstrate, the Loop Quantum Cosmology corrected theory and the classical theory are identical at leading order in the perturbative expansion quantified by the parameter $\rho_c$, which is the critical density of the quantum theory. Finally, by using the formalism of slow-roll scalar-tensor Loop Quantum Cosmology, we shall investigate how several inflationary potentials can be realized by the quantum theory, and we shall calculate directly the slow-roll indices and the corresponding observational indices. In addition, the $f(R)$ gravity frame picture is presented.
\end{abstract}

\pacs{04.50.Kd, 95.36.+x, 98.80.-k, 98.80.Cq,11.25.-w}

\maketitle



\def\pp{{\, \mid \hskip -1.5mm =}}
\def\cL{\mathcal{L}}
\def\be{\begin{equation}}
\def\ee{\end{equation}}
\def\bea{\begin{eqnarray}}
\def\eea{\end{eqnarray}}
\def\tr{\mathrm{tr}\, }
\def\nn{\nonumber \\}
\def\e{\mathrm{e}}

\section{Introduction}

The inflationary paradigm is one of the two most appealing scenarios that can successfully describe the early-time evolution of our Universe \cite{inflation2,inflation3,inflation4,reviews1}. An alternative early-time description of our Universe is offered by bouncing cosmologies, see Refs. \cite{Brandenberger:2016vhg,Cai:2015emx,Lehners:2011kr,Cai:2012va,Cai:2011tc,Cai:2013kja,Lehners:2015mra,Koehn:2015vvy,Odintsov:2015ynk,Odintsov:2015uca,Oikonomou:2015qha} for reviews and leading articles in the field. The latest Planck data \cite{Ade:2015lrj} constrained significantly the early-time era, imposing limitations on the allowed values of several observable quantities, and in effect, the number of viable cosmologies was significantly reduced. The inflationary paradigm describes the evolution of the Universe after the Big Bang, while in bouncing cosmology, the Universe contracts until a minimum radius is reached, and then it starts expanding. Moreover the process, for some models, could occur infinitely many times, although there are bouncing scenarios which can stop in a Big Rip singularity \cite{Odintsov:2015zza}. The appealing feature of bouncing cosmologies in comparison to the inflationary scenarios is that the initial singularity problem is absent in bounce cosmologies, although there exist works which combine a pre-bounce inflationary era preceding a bounce era \cite{Liu:2013kea,Piao:2003zm}.

On the other hand, Loop Quantum Cosmology (LQC) offers a fertile theoretical ground that can harbor several viable cosmological scenarios \cite{LQC1,LQC3,LQC4,LQC5,Salo:2016dsr,Xiong:2007cn,Amoros:2014tha,Cai:2014zga,deHaro:2014kxa}. Indeed, both bounce cosmology and inflationary evolutions can be realized in the context of LQC. Also the combination of modified gravity theories \cite{reviews1,reviews2,reviews4,reviews5,reviews6,reviews7} with LQC can also offer several theoretical insights and also makes possible the realization of various viable cosmologies. An appealing feature of LQC theories is that finite-time singularities are removed, at least when a single matter fluid of any sort, describes the matter content of the Universe. In principle, in the context of $f(R)$ gravity LQC, the loop quantum effects can be introduced by holonomy corrections in the classical theory, and there are infinitely many ways that this can be done \cite{Amoros:2014tha}. Also, the Einstein frame formulation of LQC $f(R)$ gravity is possible, and for a thorough examination of various theoretical features we refer the reader in Ref. \cite{Amoros:2014tha}.

In some recent works, we introduced a bottom-up reconstruction method from the observational indices for various modified gravities \cite{Odintsov:2018ggm,Odintsov:2017fnc}. Particularly, the bottom-up reconstruction method is based on fixing the functional form of the scalar-to-tensor ratio, and from it the rest of the physical quantities of the theory is reconstructed, see \cite{Odintsov:2017fnc} for the $f(R)$ gravity case, and also \cite{Odintsov:2018ggm} for mimetic and  $f(\phi)R$ theories. The purpose and main aim of this work is to generalize the bottom-up reconstruction technique of Refs. \cite{Odintsov:2018ggm,Odintsov:2017fnc} in the context of Einstein frame LQC. Particularly, we aim to realize various realistic inflationary scenarios by using a generalized bottom-up reconstruction technique in the context of LQC. To this end, we shall present the essential features of a canonical scalar theory, in the context of LQC, and in the slow-roll approximation. Here it is important to point out that we are dealing with the so-called deformed algebra approach of LQC, where, in order to introduce honolomy corrections, the Asthekar connetion is replaced by a suitable sinus function and the anomalies appearing in the
algebra of constraints are removed \cite{caitelleau, grain},  in contrast to the {\it dressed metric} approach, where the quantized perturbations evolves in a dressed
metric which encodes the quantum nature of the background (see the seminal series of papers \cite{agullo} for a detailed explanation of this new approach to LQC).

After expressing the slow-roll indices and the observational indices as functions of the $e$-foldings  number, we shall specify the functional form of the scalar-to-tensor ratio and from it we shall determine the second slow-roll index, the spectral index and the reconstructed scalar potential that can realize the given scalar-to-tensor ratio. In principle, various scenarios can be realized, but we shall be interested in realizing a specific and quite popular class of theories, the $\alpha$-attractor theories \cite{alpha1,alpha2,alpha3,alpha4,alpha5,alpha6,alpha7,alpha8,alpha9,alpha10,alpha11,alpha12,linderefs1,linder,Odintsov:2016vzz,Odintsov:2016jwr,extra1,extra2,extra3,extra4,extra5,extra6,extra7,extra8,extra9,extra10,extra11}. Many well-known inflationary scenarios, such as the Starobinsky inflation \cite{starob1,starob2} or the Higgs inflation \cite{higgs}, belong to the $\alpha$-attractor theories. Thus, we shall assume that the functional form of the scalar-to-tensor ratio is identical to the one corresponding to the $\alpha$-attractor theories, and we shall investigate which LQC-corrected scalar theory can realize such a scalar-to-tensor ratio. We shall find the analytic expressions of the spectral index and of the slow-roll indices as functions of the $e$-foldings number, and we shall provide analytic formulas that can be used easily to reproduce any given cosmological evolution. Apart from the $\alpha$-attractors example, we discuss some other characteristic examples, in order to illustrate how the method works, and we examine the possibility of realizing various cosmological scenarios. Our final task in this paper is to directly realize various inflationary scenarios in the context of LQC scalar theory, by using the formulas we developed. The difference between the two methods is that in the second case, we will express the slow-roll indices in terms of the canonical scalar field $\phi$, while in the first case, the slow-roll indices and the rest of the physical quantities will be expressed in terms of the $e$-foldings number $N$. Finally, the $f(R)$ gravity frame description of the resulting theories shall be presented too. It is worth to mention, that similar to our approach in the context of scalar-tensor cosmology, there exist various works in the literature, see for example \cite{Gao:2017uja,Lin:2015fqa,Miranda:2017juz,Fei:2017fub}.

This paper is organized as follows: In section II we present the basic equations of LQC and how can holonomy corrections be introduced. Also we present the Hamiltonian, along with the constraints, and also the modified Friedmann equation in LQC. In section III we focus on a slow-rolling LQC canonical scalar field theory, and we provide analytic expressions for the slow-roll indices and the observational indices, as functions of the scalar field. Also, the connection with the $e$-foldings number shall be given too. Furthermore we shall present the bottom-up reconstruction approach in scalar LQC theory, and we use two illustrative examples in order to show how the method works, one of which corresponds to the $\alpha$-attractor theories. Before we proceed to the examination of the examples, we describe in full detail the bottom-up reconstruction method of LQC scalar theory, and we provide analytic formulas of all the physical quantities used. Also, for all the examples we shall present, the confrontation of the resulting theory with the observational data is performed, in terms of the free parameters of the theory. Also in some cases, the classical limit of the theory is discussed. In section IV, we realized various cosmological potentials in the context of LQC cosmology scalar theory in the slow-roll approximation, and also the $f(R)$ gravity frame descriptions are found too. Finally, the conclusions follow in the end of the paper.

Before we start, let us mention that for all the considerations that are going to be made in this paper, the geometric background shall be a flat Friedmann-Robertson-Walker (FRW), with line element,
\be
\label{metricfrw} ds^2 = - dt^2 + a(t)^2
\delta_{ij}dx^idx^j,  \ee
with $a(t)$ being as usual the scale factor of the Universe. In addition, we shall assume a torsion-less, metric compatible and symmetric connection, the Levi-Civita connection. Finally, for notational convenience, we shall use a physical units system in which $\hbar=c=8\pi G=\kappa^2=1$.

\section{Essential Features of Loop Quantum Cosmology in the Einstein Frame}

In this section we briefly review the basic features of the  holonomy corrected LQC formulation (We should note here that we do not take into account inverse-volume corrections, because
at present time, the status of these corrections is not clear at all due to the fiducial-cell dependence \cite{MBojowald}). In the context of loop quantum cosmology, the spacetime has a discrete nature and this is quantified in the Hamiltonian in terms of the holonomies $h_j=e^{-\frac{i\lambda\sigma_j}{2}}$, where $\sigma_j$ are the Pauli matrices. In terms of the holonomies, the LQC Hamiltonian reads \cite{abl03,bojowald05},
\begin{equation}\label{hamiltonianlqc}
\mathrm{H}_{LQC}=-\frac{2V}{\gamma^3\lambda^3}\Sigma_{i,j,k}\epsilon^{ijk}\mathrm{Tr}[h_i(\lambda)h_j(\lambda)h_i^{-1}(\lambda)\{h_k^{-1},V\}]+\rho V\, ,
\end{equation}
where $\gamma=0.2375$
is the Barbero-Immirzi parameter, $\lambda=\sqrt{\frac{\sqrt{3}}{2}\gamma}=0.3203$ is a parameter  with dimensions of length,
whose value is equal to the square root of the minimum eigenvalue of the area operator in Loop Quantum Gravity \cite{Singh07}, $V$ is the
volume of a fixed fiducial cell (for a non compact spacetime), which for a flat FRW metric with scale factor $a$ it is equal to $V=a^3$ and finally, $\rho$ is the total effective energy density of the Universe. Also the parameter $\beta$ entering the Hamiltonian via the holonomies is the canonically conjugate variable of the volume $V$, and the Poisson bracket of these two is $\{\beta,V\}=\frac{\gamma}{2}$. By utilizing the analytic form of the holonomies and by calculating the trace appearing in the Hamiltonian (\ref{hamiltonianlqc}), the latter can be written as follows  \cite{he,dmp},
\begin{equation}\label{hamiltonianlqc2}
\mathrm{H}_{LQC}=-3V\frac{\sin^2(\lambda \beta)}{\gamma^2\lambda^2}+\rho V\, .
\end{equation}
The imposition of the Hamiltonian constraint $\mathrm{H}_{LQC}=0$ leads to the following holonomy corrected FRW equation,
\begin{equation}\label{hcqrefre1}
\frac{\sin^2(\lambda \beta)}{\gamma^2\lambda^2}=\frac{\rho}{3}\, ,
\end{equation}
and in conjunction to the Hamiltonian equation $\dot{V}=\{V,\mathrm{H}_{LQC}\}=-\frac{\gamma}{2}\frac{\partial \mathrm{H}_{LQC}}{\partial \beta}$, one obtains the following equation,
\begin{equation}\label{frwhceqn1}
H=\frac{\sin(\lambda \beta)}{\gamma \lambda}\, ,
\end{equation}
or equivalently,
\begin{equation}\label{frwhceqn2}
\beta=\frac{\arcsin (2\lambda \gamma H)}{2\lambda}\, .
\end{equation}
By substituting the obtained value of the parameter $\beta$ from Eq. (\ref{frwhceqn2}) in the Hamiltonian constraint equation (\ref{hcqrefre1}), we obtain,
\begin{equation}\label{frwhceqn3}
\frac{\sin^2(\lambda \frac{\arcsin (2\lambda \gamma H)}{2\lambda})}{\gamma^2\lambda^2}=\frac{\rho}{3}\, ,
\end{equation}
and we should note that the holonomy corrections modify the geometric sector of the Friedmann equation, so for small values of the Hubble rate $H$, the classical limit of the Friedman equation is obtained. After some algebra, the LQC Friedmann equation acquires its well-known form,
\begin{equation}\label{lqcfriedmannequation}
H^2=\frac{\rho}{3}\left( 1-\frac{\rho}{\rho_c}\right)\, ,
\end{equation}
 where the parameter $\rho_c=\frac{3}{\gamma^2\lambda^2}\cong 258$ is the so-called critical density and it is the maximum value that could reach the energy density of the
Universe. Note that  the equation (\ref{lqcfriedmannequation})
captures all the LQC quantum effects. Indeed, when the energy density of the universe is small compared with the critical one, the classical Friedmann equation $H^2=\frac{\rho}{3}$ is recovered, but for large values of the energy density the quantum effects have a strong effect on the theory, leading to a bounce, and thus,
modifying critically the cosmological evolution of the Universe.


\section{Reconstruction of Einstein Frame Loop Quantum Cosmology Inflation from the Observational Indices}

In this section,
assuming that the number of e-folds of inflation was less than $70$ in order to evade the trans-planckian problem (see \cite{martineau} for a detailed
analysis of the evolution of modes with a physical length smaller than the Planck's one), and thus obtaining the usual primordial power spectrum
for perturbations  in LQC, we shall develop a general reconstruction technique of viable inflationary cosmological evolutions by specifying the functional form of the first slow-roll index $\epsilon$, in the context of a LQC canonical scalar field.
In this theory, the canonical scalar controls the dynamics of the Universe and the LQC modified Friedmann equation is given in Eq. (\ref{lqcfriedmannequation}). The energy density and the total pressure of the canonical scalar field $\phi$ with scalar potential $V(\phi)$ is the following,
\begin{equation}\label{canonicalscalarfieldenergydensity}
\rho=\frac{\dot{\phi}^2}{2}+V(\phi),\,\,\,p=\frac{\dot{\phi}^2}{2}-V(\phi)\, ,
\end{equation}
where the ``dot'' denotes differentiation with respect to the cosmic time $t$. The equation of motion of the scalar field is equal to,
\begin{equation}\label{canonicalscalareqnofmotion}
\ddot{\phi}+3H\dot{\phi}+\frac{\partial V(\phi)}{\partial \phi}=0\, ,
\end{equation}
and by assuming that the following slow-roll conditions apply for the dynamical evolution of the scalar field, namely,
\begin{equation}\label{slowrollconditions}
\dot{\phi}^2\ll V(\phi),\,\,\,\ddot{\phi}\ll H\dot{\phi}\, ,
\end{equation}
the equation of motion for the scalar field becomes,
\begin{equation}\label{slowrolleqnofmotion}
3H\dot{\phi}\simeq -\frac{\partial V(\phi)}{\partial \phi}\, .
\end{equation}
Also by combining the modified Friedmann equation (\ref{lqcfriedmannequation}) with Eq. (\ref{canonicalscalarfieldenergydensity}) and also by taking into account the slow-roll conditions (\ref{slowrollconditions}), we obtain the following approximate relation \cite{Amoros:2014tha},
\begin{equation}\label{hasquarescalar}
H^2\simeq  \frac{V(\phi)}{3}\left( 1-\frac{V(\phi)}{\rho_c}\right)\, .
\end{equation}
The first two slow-roll indices $\epsilon$ and $\eta$ for a canonical scalar field are defined as follows,
\begin{equation}\label{slowrollindicesforcanonicalscalarfield}
\epsilon=-\frac{\dot{H}}{H^2},\,\,\,\eta=2\epsilon-\frac{\dot{\epsilon}}{2H\epsilon}\, ,
\end{equation}
so for a slow-rolling scalar field in the context of LQC, the slow-roll indices can be approximated as follows \cite{Amoros:2014tha},
\begin{equation}\label{slowrollindiceslqc}
\epsilon\simeq \frac{1}{2}\left( \frac{\frac{\partial V(\phi)}{\partial \phi}}{V(\phi)} \right)^2\frac{\left( 1-\frac{2V(\phi)}{\rho_c}\right)}{\left( 1-\frac{V(\phi)}{\rho_c}\right)^2},\,\,\,\eta \simeq \frac{1}{V(\phi)}\frac{\partial^2 V(\phi)}{\partial \phi^2}\frac{ 1}{\left( 1-\frac{V(\phi)}{\rho_c}\right)}\, .
\end{equation}
The above relations are valid, since the pivot scale leaves the Hubble radius for energy densities
of many orders less than $\rho_c$, and at that moment LQC is an small perturbation of the standard Einstein-Hilbert theory. Thus the above definition of the slow-roll parameters is approximately correct.

Finally, for a slow-rolling canonical scalar field, the spectral index of the primordial curvature perturbations $n_s$ and the scalar-to-tensor ratio $r$ as functions of the slow-roll indices $\epsilon$ and $\eta$ are equal to,
\begin{equation}\label{observationalindices}
n_s\simeq 1-6\epsilon+2\eta,\,\,\,  r\simeq 16 \epsilon\, ,
\end{equation}
which are valid at leading order in our case too, since in the long wavelength approximation, the Mukanov-Sasaki equations of LQC and standard Einstein-Hilbert gravity are the same.

Having presented the basic equations that control the scalar field dynamics in the context of LQC, we now demonstrate how the reconstruction technique from the observational indices is constructed. The starting point is the first slow-roll index $\epsilon$, which we define it to have a desirable form as function of the $e$-foldings number $N$. The particular desired form is determined essentially by the form of the scalar-to-tensor ratio we want to achieve, since $r$ is a linear function of the first slow-roll index $\epsilon$. The slow-roll indices (\ref{slowrollindiceslqc}) must be expressed as functions of the $e$-foldings number $N$, which for a LQC slow-rolling scalar field it is defined as follows \cite{Amoros:2014tha},
\begin{equation}\label{efoldingsnumber}
N=\int^{\phi_{*}}_{\phi_{end}}\frac{V(\phi) }{\frac{\partial V(\phi)}{\partial \phi}}\left( 1-\frac{V(\phi)}{\rho_c}\right)d\phi \, ,
\end{equation}
where $\phi_{*}$ and $\phi_{end}$ are the scalar field values when the pivot scale crosses the Hubble radius and   at the end of the inflationary era, respectively. The differential form of the equation (\ref{efoldingsnumber}) is,
\begin{equation}\label{differentialformofefoldings}
\left(\frac{\mathrm{d}N}{\mathrm{d}\phi}\right)^2=\frac{V(N) \left(1-\frac{V(N)}{\rho_c}\right)}{V'(N)}\, ,
\end{equation}
where the ``prime'' denotes differentiation with respect to the $e$-foldings number, and this relation will be useful in the rest of the reconstruction method. Also the first slow-roll index expressed as a function of the $e$-foldings number $N$ can be found by combining Eqs. (\ref{slowrollindiceslqc}) and (\ref{differentialformofefoldings}), and it is,

\begin{equation}\label{firstlowrollindexlqc}
\epsilon(N)\simeq \frac{V'(N)}{2V(N)}\frac{\left(1-\frac{2V(N)}{\rho_c}\right)}{\left(1-\frac{V(N)}{\rho_c}\right)}
\end{equation}
and in addition the second slow-roll index $\eta$ as a function of $N$ is,
\begin{equation}\label{secondslowrollindexlqc}
\eta \simeq \frac{1}{2}\left(\frac{V'}{V}+\frac{V''}{V'}  \right)-\frac{1}{2}\frac{V'V}{\rho_c}\frac{1}{\left(1-\frac{V}{\rho_c}\right)}\, .
\end{equation}
Then, by substituting the functional form of the first slow-roll index $\epsilon (N)$ in the first equation in Eq. (\ref{slowrollindiceslqc}), and by solving the resulting differential equation we can find the explicit form of the scalar potential as a function of the $e$-foldings number $N$, namely the function $V(N)$. After having this function at hand, we substitute it in the slow-roll index $\eta$ in Eq. (\ref{secondslowrollindexlqc}) and we can obtain its analytic form as function of $N$. Then by substituting the resulting forms of $\epsilon (N)$ and $\eta (N)$ in the observational indices (\ref{observationalindices}), we can find these as functions of $N$, so the viability can be checked directly, at leading order in $N$. It is conceivable that since $\epsilon (N)$ is appropriately chosen, the viability of the model is possibly guaranteed from the beginning of the method, but this must be checked explicitly. Finally, by using the functional form of the scalar potential $V(N)$ and substituting in Eq. (\ref{differentialformofefoldings}), we can find the functional form of the potential as a function of the scalar field, by finding the function $N(\phi )$.

Let us demonstrate explicitly how the method works by using some illustrative examples. Suppose that the first slow-roll index has the form,
\begin{equation}\label{example1alphaattractors}
\epsilon (N)=\frac{3\alpha}{4N^2}\, ,
\end{equation}
and in effect, the scalar-to-tensor ratio is,
\begin{equation}\label{scalartotensorratiolqcnew}
r\simeq \frac{12\alpha}{N^2}\, ,
\end{equation}
which is identical to the one obtained from $\alpha$-attractor models in the Einstein frame. By substituting the first slow-roll index (\ref{example1alphaattractors}) in Eq. (\ref{firstlowrollindexlqc}), the resulting differential equation can be solved analytically, so we obtain two possible scalar potentials $V(N)$
\begin{equation}\label{scalarpotentialexample1}
 V(N)=\frac{\rho_c}{2}\left( 1\pm \sqrt{1-\frac{4V_0}{\rho_c}e^{-\frac{3\alpha}{2N}}}  \right)
\end{equation}
where $V_0>0$ is an integration constant. Having the explicit form of the potential at hand, will enable us to calculate the observational indices directly. However, before proceeding to this calculation, let us investigate the classical limit of the slow-roll indices. In this case one has,
\begin{eqnarray*}
\epsilon\cong \frac{V'}{2V}\qquad \eta\cong \frac{1}{2}\left(\frac{V'}{V}+\frac{V''}{V'}  \right).
\end{eqnarray*}
Then, for the expression (\ref{example1alphaattractors}) the potential reads $V=V_0e^{-\frac{3\alpha}{2N}}$. A simple calculation leads to
$$\eta=-\frac{1}{N}+\frac{3\alpha}{2N^2}.
$$
Then, at the classical level $n_s\cong 1-\frac{2}{N}-\frac{3\alpha}{2N^2}$, which means that when one introduces holonomy corrections the result has to be
\begin{equation}\label{asx1new}
n_s\cong \left(1-\frac{2}{N}-\frac{3\alpha}{2N^2}\right)\left(1+\mathcal{O}\left(\frac{V_0}{\rho_c} \right)  \right).
\end{equation}
Indeed, such expansion is possible at the classical limit as we show later on. The choice of the minus or positive sign in the potential in Eq. (\ref{scalarpotentialexample1}) plays no essential role for the observational indices, since the resulting form of the slow-roll indices are identical for both the minus and positive sign. The sign however in Eq. (\ref{scalarpotentialexample1}) will play some important role in the determination of the functional form of the potential as a function of the canonical scalar field $\phi$. By substituting the resulting scalar potentials $V(N)$ from Eq. (\ref{scalarpotentialexample1}) in the expression appearing in Eq. (\ref{secondslowrollindexlqc}) for the second slow-roll index $\eta (N)$, the latter can be obtained and it reads,
\begin{equation}\label{secondslowrollindexexample1}
\eta =\frac{2 N \rho_c e^{\frac{3 \alpha }{2 N}}-3 \alpha  \rho_c e^{\frac{3 \alpha }{2 N}}-8 N V_0+9 \alpha  V_0}{2 N^2 \left(4 V_0-\rho_c e^{\frac{3 \alpha }{2 N}}\right)}\, .
\end{equation}
Having the slow-roll indices at hand, namely Eqs. (\ref{example1alphaattractors}) and (\ref{secondslowrollindexexample1}), we can find the analytic form of the spectral index of the primordial curvature perturbations as a function of the $e$-foldings number $N$, so by substituting Eqs. (\ref{example1alphaattractors}) and (\ref{secondslowrollindexexample1}) in Eq. (\ref{observationalindices}), the spectral index reads,
\begin{equation}\label{spectralindexexpreession1}
n_s\simeq \frac{N^2 \left(8 V_0-2 \rho_c e^{\frac{3 \alpha }{2 N}}\right)+N \left(4 \rho_c e^{\frac{3 \alpha }{2 N}}-16 V_0\right)+3 \alpha  \left(\rho_c e^{\frac{3 \alpha }{2 N}}-6 V_0\right)}{2 N^2 \left(4 V_0-\rho_c e^{\frac{3 \alpha }{2 N}}\right)}
\, .
\end{equation}
We can find the leading order behavior of the spectral index for large $N$, so by expanding Eq. (\ref{spectralindexexpreession1}) in the large-$N$ limit, we obtain the following expression,
\begin{equation}\label{leadingorderspectralindexexample1}
n_s\simeq 1-\frac{2}{N}-\frac{3 (6 \alpha  V_0-\alpha  \rho_c)}{2 N^2 (4 V_0-\rho_c)}-\frac{9 \left(\alpha ^2 \rho_c V_0\right)}{2 N^3 (4 V_0-\rho_c)^2}+\mathcal{O}(\frac{1}{N^4})\, ,
\end{equation}
and one can readily recognize that by taking into account only the first two terms, the leading order behavior of the LQC-corrected spectral index is identical to the spectral index of a canonical scalar field in the Einstein frame for the $\alpha$-attractor theories \cite{alpha1}. In addition, it is also possible to take the large $\rho_c$ limit of the spectral index (\ref{leadingorderspectralindexexample1}), so one obtains,
\begin{equation}\label{largerhoclimit}
n_s\simeq 1-\frac{2}{N}-\frac{3 \alpha }{2 N^2}-\frac{9 \alpha ^2 V_0}{2 N^3 \rho_c}\, ,
\end{equation}
and by looking the first three terms, one can readily see that these coincide with the classical limit obtained in Eq. (\ref{asx1new}).
Essentially, the third term and the higher order terms in the LQC-corrected spectral index of Eq. (\ref{leadingorderspectralindexexample1}) quantify the quantum corrections introduced by the holonomy corrections in the classical theory.

\begin{figure}[h]
\centering
\includegraphics[width=18pc]{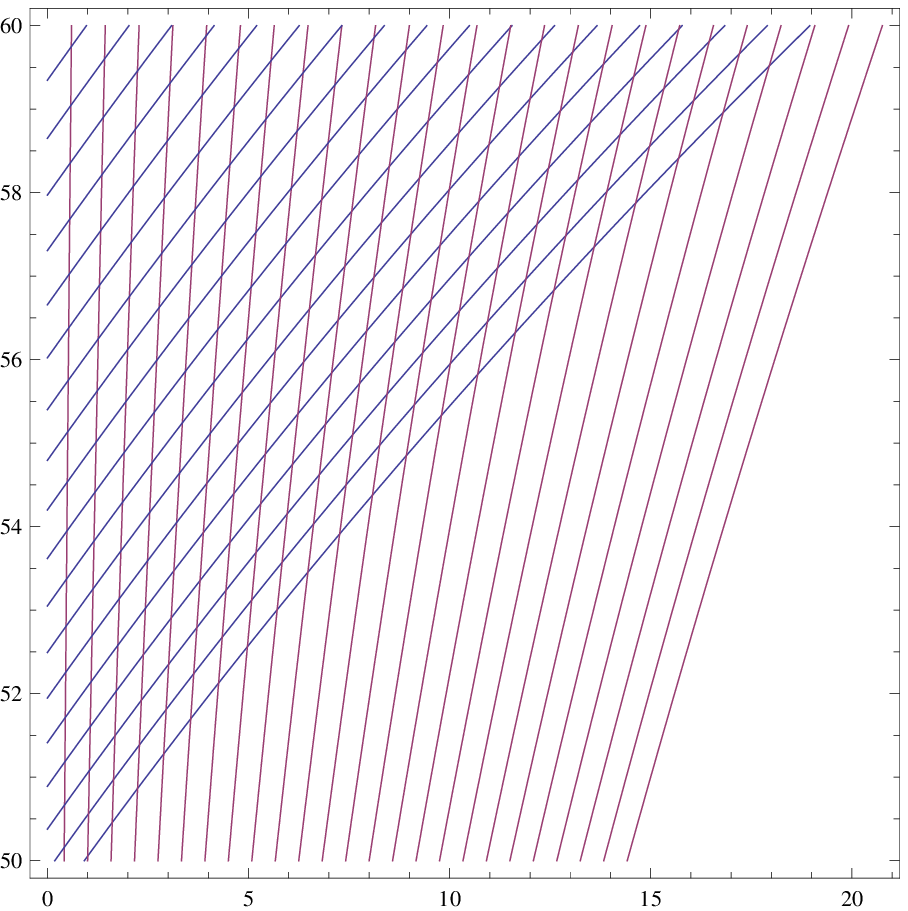}
\includegraphics[width=18pc]{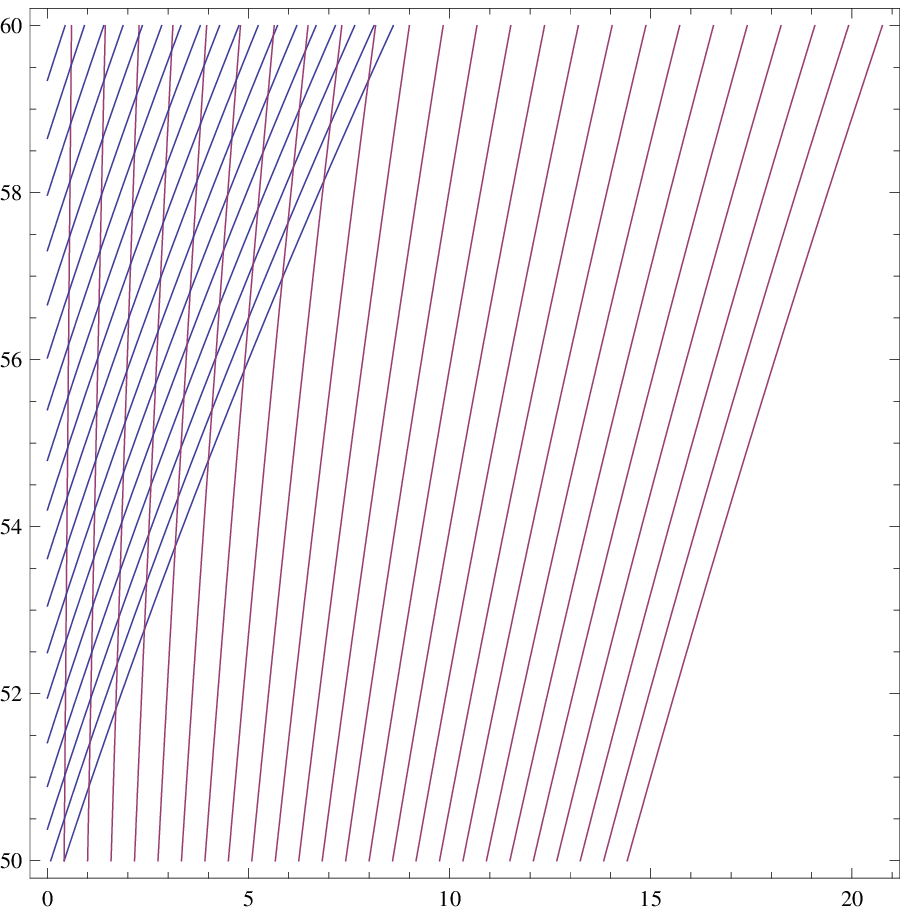}
\caption{Contour plots of the spectral index (blue curves) and of the scalar-to-tensor ratio (red curves), as functions of the $e$-foldings number $N$ and of the parameter $\alpha$, for the LQC-corrected theory (left plot) and for the classical theory (right plot), for $V_0=\mathcal{O}(10)$ and $\rho_c\simeq 258$. The curves correspond to the values for $n_s$ in the range $n_s=[0.9595,0.9693]$, for various values of the scalar-to-tensor ratio with the constraint $r<0.07$ and also the $e$-foldings number takes values in the range $N=[50,60]$ while the parameter $\alpha$ in the range $\alpha=[0,20]$.}\label{plot1}
\end{figure}

In effect, it can be seen that at leading order, the LQC-corrected spectral index is identical to the classical result for $\alpha$-attractor theories. Also, the first two terms of the quantum and of the classical expressions for the spectral index are identical, as it can be seen by directly comparing Eqs. (\ref{leadingorderspectralindexexample1}) and (\ref{largerhoclimit}). Having the LQC-corrected theory at hand, we can compare the quantum corrected theory with the classical theory, at leading-$N$ order. By taking into account the first two terms in the spectral index, the classical theory and the LQC-corrected theory are identical, so let us see how the presence of the quantum theory parameters $\rho_c$ and $V_0$ can affect the viability of the theory. After some extensive analysis of the parameter space, the resulting picture is that the presence of  $\rho_c$ and $V_0$ in the LQC-corrected theory, makes viable the inflationary evolution for larger values of $\alpha$. In order to see this explicitly, in Fig. \ref{plot1}, we present the contour plots of the spectral index (blue curves) and of the scalar-to-tensor ratio (red curves), as functions of the $e$-foldings number $N$ and of the parameter $\alpha$, for the LQC-corrected theory (left plot) and for the classical theory (right plot). We used the values $V_0=\mathcal{O}(10)$ and $\rho_c\simeq 258$. The curves correspond to the values for $n_s$ in the range $n_s=[0.9595,0.9693]$, which are the allowed ones from the latest Planck data \cite{Ade:2015lrj} and for various values of the scalar-to-tensor ratio with the constraint $r<0.07$ which is imposed by the BICEP2/Keck-Array data \cite{Array:2015xqh}. Also the $e$-foldings number takes values in the range $N=[50,60]$ while the parameter $\alpha$ in the range $\alpha=[0,20]$.
Note that in both the plots, a viable inflationary theory is depicted by the overlapping red and blue curves. As it can be seen from Fig. \ref{plot1}, the LQC-corrected theory can be come viable and compatible with the observational data for a wider range of values of the parameter $\alpha$.


Let us consider another characteristic example of cosmological evolution, so let us assume that the first slow-roll index has the following form,
\begin{equation}\label{example1alphaattractors2}
\epsilon (N)=\frac{c}{N}\, ,
\end{equation}
where $c>0$, so the scalar-to-tensor ratio reads in this case,
\begin{equation}\label{scalartotensorratiolqcnew2}
r\simeq \frac{16 c}{N}\, .
\end{equation}
By substituting the first slow-roll index (\ref{example1alphaattractors2}) in Eq. (\ref{firstlowrollindexlqc}), the resulting differential equation can be solved analytically, so the two scalar potentials $V(N)$ read in this case,
\begin{equation}\label{scalarpotentialexample12}
 V(N)=\frac{\rho_c}{2}\left(1\pm\sqrt{1-\frac{4V_0N^{2c}}{\rho_c}}\right)\, ,
\end{equation}
where $V_0>0$. By substituting the scalar potential $V(N)$ (\ref{scalarpotentialexample12}) in Eq. (\ref{secondslowrollindexlqc}) the second slow-roll index $\eta (N)$ reads,
\begin{equation}\label{secondslowrollindexexample12}
\eta =\frac{4 (3 c-1) V_0 N^{2 c}-4 c \rho_c+\rho_c}{2 N \left(4 V_0 N^{2 c}-\rho_c\right)}\, ,
\end{equation}
for both the positive or the negative sign choices in Eq. (\ref{scalarpotentialexample12}). By substituting Eqs. (\ref{example1alphaattractors2}) and (\ref{secondslowrollindexexample12}) in Eq. (\ref{observationalindices}), the resulting spectral index reads,
\begin{equation}\label{spectralindexexpreession12}
n_s\simeq \frac{-4 (3 c+1) V_0 N^{2 c}+4 V_0 N^{2 c+1}+2 c \rho_c-N \rho_c+\rho_c}{N \left(4 V_0 N^{2 c}-\rho_c\right)}
\, .
\end{equation}
We can find the leading order behavior of the spectral index  in the large-$N$ limit, so we obtain the following expression,
\begin{equation}\label{leadingorderspectralindexexample12}
n_s\simeq 1-\frac{1}{N}-\frac{2 c}{N}-\frac{4 V_0 N^{2 c}}{\rho_c}+\frac{4 V_0 N^{2 c-1}}{\rho_c}\, ,
\end{equation}
where we can see the contribution of the LQC-corrected theory quantified in the third and higher order terms. A thorough analysis of the parameter space reveals that the LQC-corrected theory is viable and compatible with both the Planck \cite{Ade:2015lrj} and the BICEP2/Keck-Array data \cite{Array:2015xqh} for a wide range of values of the variables.

\begin{figure}[h]
\centering
\includegraphics[width=18pc]{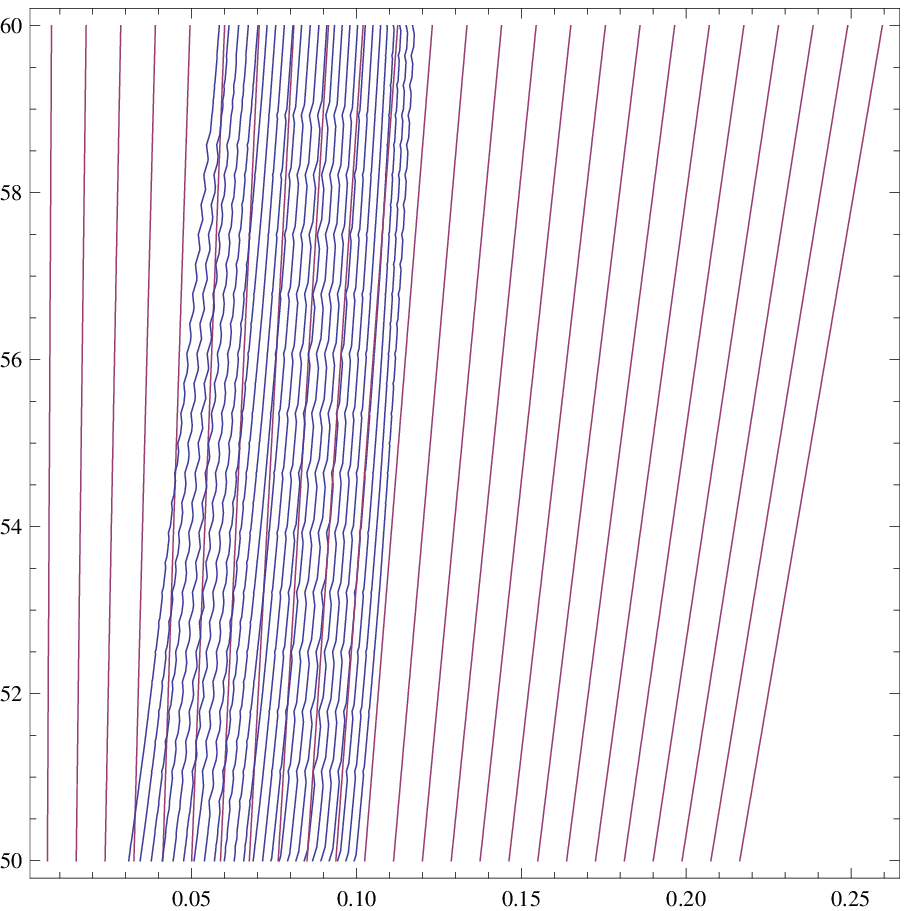}
\includegraphics[width=18pc]{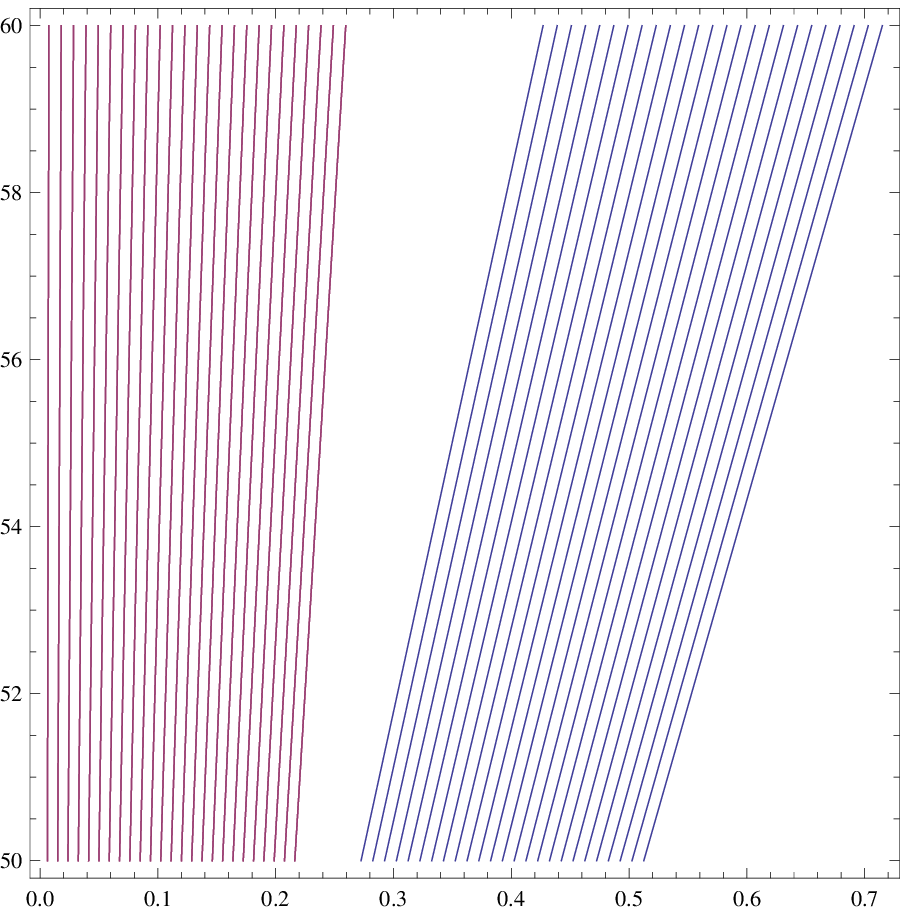}
\caption{Contour plots of the spectral index (blue curves) and of the scalar-to-tensor ratio (red curves), as functions of the $e$-foldings number $N$ and of the parameter $c$, for the LQC-corrected theory for $V_0=\mathcal{O}(0.5)$ and $\rho_c=258$ The left plot corresponds to the LQC-corrected theory and the right plot to the classical theory.}\label{plot2}
\end{figure}

In order to explicitly demonstrate this, in Fig. \ref{plot2} we present the contour plots of the spectral index (blue curves) and of the scalar-to-tensor ratio (red curves) as functions of the $e$-foldings number $N$ and of the parameter $c$, by choosing $V_0=\mathcal{O}(0.5)$ and $\rho_c=258$ (left plot). In the right plot we present the behavior of the classical spectral index and of the scalar-to-tensor ratio as a function of $N$ and of the parameter $c$. The values of the $e$-foldings number are taken in the range $N=[50,60]$ and of $c$ in the range $c=[0,1]$. Also for the contour plots, the spectral index is assumed to take values in the range $n_s=[0.9595,0.9693]$ which is compatible with the Planck data \cite{Ade:2015lrj}, while the scalar-to-tensor ratio takes values in the range $r=[0,0.07]$, which compatible with the BICEP2/Keck-Array data \cite{Array:2015xqh}.
As it can be seen in Fig. \ref{plot2}, the regions of overlapping blue and red curves indicate the regions of viability of the theory, so it can be seen that this can be achieved for a large range of the parameters, and only for the LQC-corrected theory.

\section{Specific Potentials with Einstein Frame Loop Quantum Cosmology and $f(R)$ Gravity Picture}

In this section we shall realize directly the slow-roll inflationary era of various well-known scalar potentials, and also we shall find the corresponding $f(R)$ gravity frame description. In the Jordan frame, for the flat FRW metric of Eq. (\ref{metricfrw}) and in the absence of matter component, the action is given by,
 \begin{equation}\label{JF-action}
S_{JF}=\int   \frac{1}{2}f(R)a^3dt,
 \end{equation}
The passage to the Einstein frame  is found by performing the conformal transformation
 $\bar{a}=\sqrt{F(R)}a$,
and $d\bar{t}=\sqrt{F(R)}dt$,
where $F(R)=f_R(R)$.
Then, in this frame the action is given by
 \begin{equation}\label{EF}
  S_{EF}=\int \left(\frac{1}{2}\bar{R}+{\mathcal L}_{matt}\right){\bar a}^3 d\bar{t},
 \end{equation}
where ${\mathcal L}_{matt}$ is  the  matter Lagrangian, which for an scalar field is the pressure,  is given by
 \begin{eqnarray}
  {\mathcal L}_{matt}=p
  =\frac{\dot{\phi}^2}{2}
  -V(\phi).
  \end{eqnarray}
where,
\begin{eqnarray}\label{A}
{\phi}\equiv \sqrt{\frac{3}{2}}\ln\left( F(R)\right), \quad \mbox{and}  \quad V(\phi)\equiv \frac{RF(R)-f(R)}{2 F^2(R)}.
\end{eqnarray}

Now, at the background level, it is formally easy to introduce holonomy correction effects. On can do it as usual, considering
the Hamiltonian  ${\mathcal H}=-3\bar{H}^2+\rho$ and performing the replacement $\bar{H}\rightarrow \frac{\sin(\gamma\lambda\bar H)}{\gamma\lambda}$ \cite{LQC1}, where, once again, $\gamma$ is the Barbero-Immirzi parameter and $\lambda$ is the
minimum eigenvalue of the area operator in Loop Quantum Gravity. Equivalently, in the context of the Lagrangian formalism,  replacing $\bar{R}$ by $\bar{R}+ g(\bar{\mathcal S})$ where $ \bar{\mathcal S}$ is an scalar that
for our metric satisfies $\bar{\mathcal S}=-6\bar{H}^2$ and $g$ is bi-valued quantity given by \cite{helling,ds09,haro12},
\begin{eqnarray}
 g(\bar{\mathcal S})=\rho_c\left(1- \sqrt{1-s^2}- s\arcsin s  \right)-\bar{\mathcal S},
 \end{eqnarray}
 where
 $s\equiv \sqrt{-\frac{2\bar{\mathcal S} }{\rho_c}},$
 and
  the sign of the square root has been chosen to be positive (respectively negative) in the lower  (respectively upper) branch  and
$ \arcsin s\equiv \int_0^s \frac{1}{\sqrt{1-{\bar s}^2}}  {d\bar{s}}$ in the lower branch, whereas  $\arcsin s\equiv \int_0^s \frac{1}{\sqrt{1-{\bar s}^2}}  {d\bar{s}}+\pi$, in the upper one, with the same criteria for the sign of the square root.

 The scalar ${\mathcal S}$ could be the  torsion obtained using the Weitzenb\"ok connection \cite{w}, the extrinsic curvature \cite{ha17},  i.e. given a  field $\varphi$ satisfying $-\nabla_{\mu}\phi\nabla^{\mu} \phi=1$ one can consider
 $\bar{\mathcal S}\equiv   (\nabla^{\nu}\nabla^{\mu}\phi)(\nabla_{\nu}\nabla_{\mu}\phi)    -( \nabla_{\mu}\nabla^{\mu}\phi)^2$
 or one obtained from the  Carminati-McLenaghan invariants \cite{cm,hp18}, i.e.,
$\bar{\mathcal{S}}\equiv -6\frac{{\mathcal {\bar R}}_3}{{\mathcal {\bar R}}_2}-\frac{\bar R}{2} $, where,
 \begin{eqnarray}
 {\mathcal{\bar R}}_2\equiv \frac{1}{4}{\mathcal {\bar R}}_{\mu}^{\nu}{\mathcal {\bar R}}_{\nu}^{\mu}, \quad \mbox{ and } \quad
 {\mathcal {\bar R}}_3\equiv -\frac{1}{8}{\mathcal {\bar R}}_{\mu}^{\nu}{\mathcal {\bar R}}_{\gamma}^{\mu}{\mathcal {\bar R}}^{\gamma}_{\nu},
\end{eqnarray}
with ${\mathcal {\bar R}}_{\mu \nu}$ being the trace-less Ricci tensor.


In effect, the difference between the quantum and the classical Einstein frame theory is that the Friedmann equation, for synchronous co-moving observers,  is given in Eq. (\ref{lqcfriedmannequation}). The difference in comparison with the standard classical Friedmann equation $\bar{H}^2=\frac{\rho}{3}$, is that the usual one represents in the plane $(\bar{H}, \rho )$ a parabola which is an unbounded curve, and thus, allowing the Big Bang and various types of rip singularities. However, the LQC modified Friedman equation is represented by an ellipse, which is closed and bounded, and consequently, crushing type singularities are removed, because the Hubble parameter and the energy density are always bounded.


\

 On the other hand,
from equation (\ref{A}) we obtain the differential equation,
\begin{equation}
F(R)R= \sqrt{6}\frac{\mathrm{d}}{\mathrm{d}\phi}\left(e^{2\sqrt{\frac{2}{3}}\phi} V(\phi)  \right),
\end{equation}
which together with $\phi=\sqrt{\frac{3}{2}}\ln F(R)$ leads to the algebraic equation,
\begin{equation}
\frac{R}{F}=\sqrt{6}\left[
2\sqrt{\frac{2}{3}}V\left( \sqrt{\frac{3}{2}}\ln F \right)+V_{\phi}\left( \sqrt{\frac{3}{2}}\ln F \right)
\right],
\end{equation}
which, in principle, could be solved for any potential $V$ and allows us to reconstruct the corresponding $f(R)$ theory.

As a simple example we consider the potential $V(\phi )=V_0\left(\cosh \left(\sqrt{\frac{2}{3}}\phi \right)-1\right)$, which leads to the equation,
\begin{equation}
3F(R)^2-4F(R)+1-\frac{R}{V_0}=0,
\end{equation}
the solution of which is,
\begin{equation}
F(R)=\frac{1}{3}\left(2\pm\sqrt{1+\frac{3R}{V_0}}                       \right),
\end{equation}
but the only physical solution is $F(R)=\frac{1}{3}\left(2+\sqrt{1+\frac{3R}{V_0}}   \right)$,  because for small values of $R$ one has $F(R)\cong 1$, and thus,  $f(R)\cong R$ recovering GR.


\

Let us focus on the realization of some specific potentials in LQC. We consider the model $f(R)=R+aR^n$, with $1<n\leq 2$, with leads to the potential,
\begin{equation}
V(\phi)=\frac{1}{2}(n-1)a\left(\frac{1}{na}  \right)^{\frac{n}{n-1}}e^{\frac{2-n}{n-1}\sqrt{\frac{2}{3}}\phi }
\left(1- e^{-\sqrt{\frac{2}{3}}\phi}\right)^{\frac{n}{n-1}},
\end{equation}
which for the particular case $n=2$ leads to the well-known potential,
\begin{equation}\label{example1}
V(\phi)=\frac{1}{8a}\left( 1-e^{-\sqrt{\frac{2}{3}}\phi }  \right)^2.
\end{equation}
In LQC the dynamical evolution implied by this potential is very simple. Since it has only one global minimum at $\phi=0$, the scalar field is at early times, during the contracting phase,
at the minimum and starts to oscillate  growing its energy density up to the critical energy density $\rho_c$. At that point the Universe bounces and enters in the contracting phase, where the scalar field slow-rolls in the potential producing an inflationary era. Finally, when it arrives to the minimum, it starts to oscillate releasing its energy and creating enough particles to reheat the Universe. Note that, this behavior agrees with the one depicted in \cite{martineau1} for the  isotropic case.
Effectively, in that paper it has been showed that taken initial conditions at very late times in the contracting phase, the scalar field  oscillates approximately as an harmonic oscillator, and for a wide range of initial conditions the number of e-folds is smaller than $70$, and consequently, evading the trans-planckian problem.

To study inflation, we first consider the case $n=2$, i.e.,  the well known potential coming form $R^2$ gravity, which  in the Einstein Frame, is given by,
(\ref{example1}).
By using introducing the parameter $V_0=\frac{1}{8a}$, the definition of the slow-roll indices and of the $e$-foldings number which we developed in the previous section, namely Eqs. (\ref{slowrollindiceslqc}) and (\ref{efoldingsnumber}), for the potential (\ref{example1}), for large field values we obtain,
\begin{equation}
\epsilon_*\cong \frac{4}{3}e^{-\sqrt{\frac{8}{3}}\phi_*}\frac{1-\frac{2V_0}{\rho_c}}{\left(1-\frac{V_0}{\rho_c}\right)^2}, \quad
\eta_*\cong -\frac{4}{3}e^{-\sqrt{\frac{2}{3}}\phi_*}\frac{1}{1-\frac{V_0}{\rho_c}},
\end{equation}
and the number of $e$-foldings is given by,
\begin{equation}
N\cong \frac{3}{4}e^{\sqrt{\frac{2}{3}}\phi_*   }\left(1-\frac{V_0}{\rho_c}\right).
\end{equation}
By combining the above, we obtain,
\begin{equation}
\epsilon_*\cong \frac{3}{4N^2}\left(1-\frac{2V_0}{\rho_c}\right),\quad \eta_*\cong -\frac{1}{N}.
\end{equation}
In effect, the observational indices (\ref{observationalindices}) read,
\begin{equation}
n_s=1-\frac{2}{N}, \qquad r=\frac{12\alpha}{N^2},
\end{equation}
where we have defined $\alpha=1-\frac{2V_0}{\rho_c}$. Hence, in the classical limit $\rho_c\gg V_0$, one acquires the  observational indices of the well-known Starobinsky model \cite{starob1}.

If we consider the potential,
\begin{equation}\label{example2}
V=V_0\left(1-e^{-\sqrt{\frac{2}{3}}\phi  }\right)^{2n}\, ,
\end{equation}
a simple calculation leads to,
\begin{equation}
\epsilon_* \cong \frac{3}{4N^2}\left(1-\frac{2V_0}{\rho_c}\right),\quad \eta_*\cong -\frac{1}{N}.
\end{equation}
and thus,
\begin{equation}
n_s=1-\frac{2}{N}, \qquad r=\frac{12\alpha}{N^2},
\end{equation}
and once again  we defined $\alpha=1-\frac{2V_0}{\rho_c}$. In this case too, at leading order and for $\rho_c\gg V_0$, one gets asymptotically the Starobinsky model.


Let us now consider the potential,
\begin{equation}\label{example3}
V=V_0\tanh^{2n} \left(\frac{\phi}{\sqrt{6}}\right)\, ,
\end{equation}
in which case the first slow-roll index reads,
\begin{equation}
\epsilon_*
=\frac{4n^2}{3}\frac{1}{\sinh^2 \left(\sqrt{\frac{2}{3}}\phi_* \right)}\frac{1-\frac{2V_0}{\rho_c}}{\left(1-\frac{V_0}{\rho_c}\right)^2}
\cong \frac{16n^2}{3}e^{-\sqrt{\frac{8}{3}}\phi_*}\frac{1-\frac{2V_0}{\rho_c}}{\left(1-\frac{V_0}{\rho_c}\right)^2},
\end{equation}
and accordingly, the second slow-roll index reads,
\begin{equation}
\eta_*\cong -\frac{2n}{3}\frac{1}{\cosh^2\left(\phi_*  \right)}\frac{1}{1-\frac{V_0}{\rho_c}}
 \cong -\frac{8n}{3}e^{-\sqrt{\frac{2}{3}}\phi_*}\frac{1}{1-\frac{V_0}{\rho_c}}.
\end{equation}
Also, the $e$-foldings number as a function of the scalar field is,
\begin{equation}
N\cong \frac{3}{8n}e^{\sqrt{\frac{2}{3}}\phi_*}\left(1-\frac{V_0}{\rho_c}\right)\, .
\end{equation}
In effect, we have,
\begin{equation}
\epsilon_*\cong \frac{3}{4N^2}\left(1-\frac{2V_0}{\rho_c}\right),\quad \eta_*\cong -\frac{1}{N},
\end{equation}
and therefore the observational indices are,
\begin{equation}
n_s=1-\frac{2}{N}, \qquad r=\frac{12\alpha}{N^2},
\end{equation}
where $\alpha=1-\frac{2V_0}{\rho_c}$. In this case the reconstruction method for the Jordan frame leads to the following algebraic equation,
\begin{equation}
R=2 V_0F\left(\frac{F-1}{F+1}  \right)^{2n-1}\left[-n\left(\frac{F-1}{F+1}\right)^{2}+2 \frac{F-1}{F+1} +n \right].
\end{equation}
In principle it is hard to solve the above algebraic equation analytically, however it is important to note that for $R=0$, a solution is $F=1$ what means that for small values of $R$ one has $F\cong 1$ and thus $f(R)\cong R$ recovering the standard Einstein-Hilbert gravity. Therefore, in the inflationary regime the $f(R)$ theory for both potentials
$V=V_0\left(1-e^{-\sqrt{\frac{2}{3}}{\phi}  }\right)^{2n} $ and $V=V_0\tanh^{2n} \left(\frac{\phi}{\sqrt{6}}\right)$
 is the same,  given by $f(R)=\frac{R^2}{8V_0}$.

As a final example, let us consider the potential $V=V_0e^{-\sqrt{\frac{2}{q}}{\phi}}$, and some variant forms of it later on. When the correspondence with the $f(R)$ gravity frame is considered, one obtains the following algebraic equation,
\begin{eqnarray}
\frac{R}{F(R)}=2V_0\left( 2-\sqrt{\frac{3}{q}} \right)F(R)^{-\sqrt{\frac{3}{q}} }\, ,
\end{eqnarray}
which has the following solution,
\begin{eqnarray}
F(R)=\left(  \frac{R}{2V_0 \left( 2-\sqrt{\frac{3}{q}} \right)  }     \right)^{\frac{1}{1-\sqrt{\frac{3}{q}}} },
\end{eqnarray}
which is a rather unphysical Jordan frame theory, because for small values of $R$, the standard Einstein-Hilbert solution is not recovered.

We have seen that for potentials like the one coming from $R^2$ gravity, the LQC effects reduce the ratio of tensor to scalar perturbations which is a good feature because observational data leads to a very small value of this quantity
($r\leq 0.12$). However, this does not happen for all the models, as an example
we consider the an extension of the potential $V=V_1e^{-\sqrt{\frac{2}{q}}{\phi}}$, which as we have showed just above, leads to an unphysical $f(R)$ theory. However, in order for this theory to be viable, it has to be matched with another potential, which for small values of the the field leads to $F(R)\cong 1$.  For example, it could be matched with
$V=V_0\tanh^{2n} \left(\frac{\phi}{\sqrt{6}}\right)$. So, the resulting potential is,
\begin{eqnarray}
V=\left\{\begin{array}{ccc}
V_0\tanh^{2n} \left(\frac{\phi}{\sqrt{6}}\right)&\mbox{for}& \phi\leq \bar{\phi}\\
V_1e^{-\sqrt{\frac{2}{q}}{\phi}}&\mbox{for}& \phi\geq \bar{\phi}>0,
\end{array}\right.
\end{eqnarray}
where $V_1=V_0\tanh^{2n} \left(\frac{\bar\phi}{\sqrt{6}}\right)e^{\sqrt{\frac{2}{q}}{\bar\phi}}$. In this case we obtain,
\begin{eqnarray}
\epsilon_*=\frac{1}{q}\frac{1-\frac{2V_*}{\rho_c}}{\left(1-\frac{V_*}{\rho_c}\right)^2}
\quad \eta_*=\frac{2}{q}\frac{1}{1-\frac{V_*}{\rho_c}},
\end{eqnarray}
where $V_*$ is $V_*=V_0e^{-\frac{2}{q}N}$. Since the number of $e$-foldings is approximately $N\cong \sqrt{\frac{q}{2}}\phi_*$, one has
$V_*=V_0e^{-\frac{2}{q}N}\cong V_0e^{-\frac{120}{q}}$, for $60$ $e$-foldings. Then, for LQC the critical energy density $\rho_c$ is always greater than $V$ for the allowed values of the scalar field, meaning that
$\rho_c \geq  V_0e^{-\frac{120}{q}}$, and thus, for values of the parameter $q$ of the order $1$, one has $\rho_c\gg V_*$. Consequently, the slow roll parameters are in this case,
\begin{eqnarray}
\epsilon_*=\frac{1}{q} \quad \mbox{and}\quad
\quad \eta_*=\frac{2}{q},
\end{eqnarray}
and coincide with the values obtained in the standard Einstein-Hilbert theory.

A final remark about the dynamics is in order: All the potentials we studied have only a global minimum, in effect, for the LQC theory, at very early times when the Universe is in the contracting
phase, the scalar field is at the minimum of the potential. Hence, due to the fact that in the contracting phase the energy density of the scalar field increases, it starts to oscillate around the minimum of the potential, releasing energy. During the oscillating phase, when the energy density reaches the critical value $\rho_c$, the Universe bounces off, entering in the expanding phase, where the field starts to slow-roll down to its minimum. In effect, the Universe enters in the inflationary epoch that ends when the scalar field starts to oscillate around the minimum,  releasing its energy to produce enough particles coupled with the field and this provides a reheating mechanism for the Universe.

\section{Conclusions}

In this work we extended the bottom-up reconstruction approach from the observational indices of Refs. \cite{Odintsov:2018ggm,Odintsov:2017fnc} in the case of a LQC scalar field, in the slow-roll approximation. In order for the method to work, the functional form of the scalar-to-tensor ratio as a function of the $e$-foldings number must be given, and from it, all the observable quantities and all the physical quantities of the theory can be determined analytically in terms of the $e$-foldings number. Also, in some cases, if the function $\phi (N)$ can be inverted, the scalar potential can be found analytically in terms of the scalar field, and in the contrary case, some approximation must be used. Also, we provided analytic and detailed formulas for the slow-roll indices and the observational indices as functions of the $e$-foldings number, and in principle any arbitrary cosmological evolution can be realized. We used two simple examples, one of which corresponds classically to the $\alpha$-attractors scalar theories. For this example, we investigated the essential features of the LQC-corrected theory and we provided analytic formulas for the slow-roll indices and for the observational indices. Also we investigated the classical limit of the theory, and we showed explicitly that the classical and the quantum theory coincide at leading order in the inverse $\rho_c$-expansion, in the case $\rho_c\gg \rho$. It is vital to note that in order for the method to work, the slow-roll approximation must hold true at all stages of the reconstruction technique. Hence, scalar theories for which one of the two slow-roll indices is of the order $\mathcal{O}(1)$ cannot be realized by using this technique. In the second part of this work, we used the formalism of LQC scalar theory in order to find directly the LQC-corrected version of the theory. In principle, the two-approaches should coincide, and they do to some extent, however some discrepancies can be found, due to the fact that for the shake of analytic results, the approximations made in the second approach may make the resulting theory to be different at higher order, in comparison to the first method results. However, at leading order these two should coincide.

Finally, it would be interesting to extend the bottom-up reconstruction method we developed in this paper, to the case of LQC-corrected $f(R)$ gravity, however, finding the LQC-corrected slow-roll indices in the Jordan frame is a challenging task, that must be addressed carefully in another work focused exactly on this issue. We hope to address this challenging task in a future work.

\section*{Acknowledgments}

This work is supported by MINECO (Spain), FIS2016-76363-P (S.D.O) and MTM2017-84214-C2-1-P  (J.d H), also by project SGR247 (AGAUR, Catalonia), 2017 (S.D.O and J. d H).


\begin{thebibliography}{99}


\bibitem{inflation2} D.~S.~Gorbunov and V.~A.~Rubakov,
``Introduction to the theory of the early universe: Cosmological
perturbations and inflationary theory,'' Hackensack, USA: World
Scientific (2011) 489 p;
%


\bibitem{inflation3}A.~Linde,
arXiv:1402.0526 [hep-th];


\bibitem{inflation4}D.~H.~Lyth and A.~Riotto,
Phys.\ Rept.\  {\bf 314} (1999) 1 [hep-ph/9807278].


\bibitem{reviews1}
 S.~Nojiri, S.~D.~Odintsov and V.~K.~Oikonomou,
  Phys.\ Rept.\  {\bf 692} (2017) 1
  doi:10.1016/j.physrep.2017.06.001
  [arXiv:1705.11098 [gr-qc]].


\bibitem{Brandenberger:2016vhg}
  R.~Brandenberger and P.~Peter,
  Found.\ Phys.\  {\bf 47} (2017) no.6,  797
  doi:10.1007/s10701-016-0057-0
  [arXiv:1603.05834 [hep-th]].





\bibitem{Cai:2015emx}
  Y.~F.~Cai, S.~Capozziello, M.~De Laurentis and E.~N.~Saridakis,
  Rept.\ Prog.\ Phys.\  {\bf 79} (2016) no.10,  106901
  doi:10.1088/0034-4885/79/10/106901
  [arXiv:1511.07586 [gr-qc]].




\bibitem{Lehners:2011kr}
  J.~L.~Lehners,
  Class.\ Quant.\ Grav.\  {\bf 28} (2011) 204004
  doi:10.1088/0264-9381/28/20/204004
  [arXiv:1106.0172 [hep-th]].








\bibitem{Cai:2012va}
  Y.~F.~Cai, D.~A.~Easson and R.~Brandenberger,
  JCAP {\bf 1208} (2012) 020
  doi:10.1088/1475-7516/2012/08/020
  [arXiv:1206.2382 [hep-th]].


\bibitem{Cai:2011tc}
  Y.~F.~Cai, S.~H.~Chen, J.~B.~Dent, S.~Dutta and E.~N.~Saridakis,
  Class.\ Quant.\ Grav.\  {\bf 28} (2011) 215011
  doi:10.1088/0264-9381/28/21/215011
  [arXiv:1104.4349 [astro-ph.CO]].


\bibitem{Cai:2013kja}
  Y.~F.~Cai, E.~McDonough, F.~Duplessis and R.~H.~Brandenberger,
  JCAP {\bf 1310} (2013) 024
  doi:10.1088/1475-7516/2013/10/024
  [arXiv:1305.5259 [hep-th]].



\bibitem{Lehners:2015mra}
  J.~L.~Lehners and E.~Wilson-Ewing,
  JCAP {\bf 1510} (2015) no.10,  038
  doi:10.1088/1475-7516/2015/10/038
  [arXiv:1507.08112 [astro-ph.CO]].



\bibitem{Koehn:2015vvy}
  M.~Koehn, J.~L.~Lehners and B.~Ovrut,
  Phys.\ Rev.\ D {\bf 93} (2016) no.10,  103501
  doi:10.1103/PhysRevD.93.103501
  [arXiv:1512.03807 [hep-th]].


\bibitem{Odintsov:2015ynk}
  S.~D.~Odintsov and V.~K.~Oikonomou,
  Int.\ J.\ Mod.\ Phys.\ D {\bf 26} (2017) no.08,  1750085
  doi:10.1142/S0218271817500857
  [arXiv:1512.04787 [gr-qc]].


\bibitem{Odintsov:2015uca}
  S.~D.~Odintsov, V.~K.~Oikonomou and E.~N.~Saridakis,
  Annals Phys.\  {\bf 363} (2015) 141
  doi:10.1016/j.aop.2015.08.021
  [arXiv:1501.06591 [gr-qc]].


\bibitem{Oikonomou:2015qha}
  V.~K.~Oikonomou,
  Phys.\ Rev.\ D {\bf 92} (2015) no.12,  124027
  doi:10.1103/PhysRevD.92.124027
  [arXiv:1509.05827 [gr-qc]].





\bibitem{Ade:2015lrj}
  P.~A.~R.~Ade {\it et al.} [Planck Collaboration],
  Astron.\ Astrophys.\  {\bf 594} (2016) A20
  doi:10.1051/0004-6361/201525898
  [arXiv:1502.02114 [astro-ph.CO]].







\bibitem{Odintsov:2015zza}
  S.~D.~Odintsov and V.~K.~Oikonomou,
  Phys.\ Rev.\ D {\bf 92} (2015) no.2,  024016
  doi:10.1103/PhysRevD.92.024016
  [arXiv:1504.06866 [gr-qc]].




\bibitem{Liu:2013kea}
  Z.~G.~Liu, Z.~K.~Guo and Y.~S.~Piao,
  Phys.\ Rev.\ D {\bf 88} (2013) 063539
  doi:10.1103/PhysRevD.88.063539
  [arXiv:1304.6527 [astro-ph.CO]].


\bibitem{Piao:2003zm}
  Y.~S.~Piao, B.~Feng and X.~m.~Zhang,
  Phys.\ Rev.\ D {\bf 69} (2004) 103520
  doi:10.1103/PhysRevD.69.103520
  [hep-th/0310206].




\bibitem{LQC1}


A.~Ashtekar and P.~Singh,
Class.\ Quant.\ Grav.\  {\bf 28} (2011) 213001
[arXiv:1108.0893 [gr-qc]]



\bibitem{LQC3} A.~Ashtekar, T.~Pawlowski and P.~Singh,
  Phys.\ Rev.\ Lett.\  {\bf 96} (2006) 141301
  [gr-qc/0602086].


\bibitem{LQC4} A.~Ashtekar, T.~Pawlowski and P.~Singh,
  Phys.\ Rev.\ D {\bf 73} (2006) 124038
  [gr-qc/0604013].


\bibitem{LQC5}  A.~Ashtekar, T.~Pawlowski and P.~Singh,
  Phys.\ Rev.\ D {\bf 74} (2006) 084003
  [gr-qc/0607039].







\bibitem{Salo:2016dsr}
  L.~Areste Salo, J.~Amoros and J.~de Haro,
  Class.\ Quant.\ Grav.\  {\bf 34} (2017) no.23,  235001
  doi:10.1088/1361-6382/aa9311
  [arXiv:1612.05480 [gr-qc]].



\bibitem{Xiong:2007cn}
  H.~H.~Xiong, T.~Qiu, Y.~F.~Cai and X.~Zhang,
  Mod.\ Phys.\ Lett.\ A {\bf 24} (2009) 1237
  doi:10.1142/S0217732309030667
  [arXiv:0711.4469 [hep-th]].

\bibitem{Amoros:2014tha}
  J.~Amoros, J.~de Haro and S.~D.~Odintsov,
  Phys.\ Rev.\ D {\bf 89} (2014) no.10,  104010
  doi:10.1103/PhysRevD.89.104010
  [arXiv:1402.3071 [gr-qc]].


\bibitem{Cai:2014zga}
  Y.~F.~Cai and E.~Wilson-Ewing,
  JCAP {\bf 1403} (2014) 026
  doi:10.1088/1475-7516/2014/03/026
  [arXiv:1402.3009 [gr-qc]].


\bibitem{deHaro:2014kxa}
  J.~de Haro and J.~Amoros,
  JCAP {\bf 1408} (2014) 025
  doi:10.1088/1475-7516/2014/08/025
  [arXiv:1403.6396 [gr-qc]].




\bibitem{reviews2}

S. Nojiri, S.D. Odintsov,
   Phys.\ Rept.\  {\bf 505}, 59 (2011); \\
S. Nojiri, S.D. Odintsov,
  eConf {\bf C0602061}, 06 (2006)
  [Int.\ J.\ Geom.\ Meth.\ Mod.\ Phys.\  {\bf 4}, 115 (2007)].


   \bibitem{reviews4}
 S. Capozziello, M. De Laurentis,
   Phys.\ Rept.\  {\bf 509}, 167 (2011);


\bibitem{reviews5} V.~Faraoni and S.~Capozziello,
  Fundam.\ Theor.\ Phys.\  {\bf 170} (2010).
  doi:10.1007/978-94-007-0165-6


\bibitem{reviews6}

A.~de la Cruz-Dombriz and D.~Saez-Gomez,
  Entropy {\bf 14} (2012) 1717
  doi:10.3390/e14091717
  [arXiv:1207.2663 [gr-qc]].



\bibitem{reviews7}

  G.~J.~Olmo,
  Int.\ J.\ Mod.\ Phys.\ D {\bf 20} (2011) 413
  doi:10.1142/S0218271811018925
  [arXiv:1101.3864 [gr-qc]].






\bibitem{Odintsov:2018ggm}
  S.~D.~Odintsov and V.~K.~Oikonomou,
  arXiv:1801.10529 [gr-qc].


\bibitem{Odintsov:2017fnc}
  S.~D.~Odintsov and V.~K.~Oikonomou,
  Annals Phys.\  {\bf 388} (2018) 267
  doi:10.1016/j.aop.2017.11.026
  [arXiv:1710.01226 [gr-qc]].





\bibitem{caitelleau}
T. Cailleteau, J. Mielczarek, A. Barrau and J. Grain,
    Class.Quantum Grav. {\bf 29}, 095010 (2012)     [arXiv:1111.3535].
\bibitem{grain}
T. Cailleteau, A. Barrau, J. Grain and F. Vidotto,
    Phys. Rev. {\bf D 86}, 087301 (2012)    [arXiv:1206.6736].
\bibitem{agullo}
 I. Agull\'o, A. Ashtekar and W. Nelson,
    Phys. Rev. Lett. {\bf 109}, 251301 (2012)   [arXiv:1209.1609];
        Phys. Rev. {\bf D87}, 043507 (2013)     [arXiv:1211.1354];
            Class. Quant. Grav. {\bf 30}, 085014 (2013)     [arXiv:1302.0254].



\bibitem{alpha1} R.~Kallosh and A.~Linde,
  JCAP {\bf 1307} (2013) 002
  [arXiv:1306.5220 [hep-th]].


\bibitem{alpha2} S.~Ferrara, R.~Kallosh, A.~Linde and M.~Porrati,
  Phys.\ Rev.\ D {\bf 88} (2013) no.8,  085038
  [arXiv:1307.7696 [hep-th]].


\bibitem{alpha3}R.~Kallosh, A.~Linde and D.~Roest,
  JHEP {\bf 1311} (2013) 198
  [arXiv:1311.0472 [hep-th]].






\bibitem{alpha4} M.~Galante, R.~Kallosh, A.~Linde and D.~Roest,
  Phys.\ Rev.\ Lett.\  {\bf 114} (2015) no.14,  141302
  [arXiv:1412.3797 [hep-th]].





\bibitem{alpha5} S.~Cecotti and R.~Kallosh,
  JHEP {\bf 1405} (2014) 114
  [arXiv:1403.2932 [hep-th]].



\bibitem{alpha6} J.~J.~M.~Carrasco, R.~Kallosh and A.~Linde,
  JHEP {\bf 1510} (2015) 147
  [arXiv:1506.01708 [hep-th]].



\bibitem{alpha7}    A.~Linde,
  JCAP {\bf 1505} (2015) 003
  doi:10.1088/1475-7516/2015/05/003
  [arXiv:1504.00663 [hep-th]].




\bibitem{alpha8} D.~Roest and M.~Scalisi,
  Phys.\ Rev.\ D {\bf 92} (2015) 043525
  doi:10.1103/PhysRevD.92.043525
  [arXiv:1503.07909 [hep-th]].




\bibitem{alpha9}  R.~Kallosh, A.~Linde and D.~Roest,
  JHEP {\bf 1408} (2014) 052
  doi:10.1007/JHEP08(2014)052
  [arXiv:1405.3646 [hep-th]].



\bibitem{alpha10} Z.~Yi and Y.~Gong,
  arXiv:1608.05922 [gr-qc].

  \bibitem{alpha11}

   Q.~Gao and Y.~Gong,
  arXiv:1703.02220 [gr-qc].


  \bibitem{alpha12}  Z.~Yi and Y.~Gong,
  arXiv:1709.04252 [gr-qc].



\bibitem{linderefs1} R.~Kallosh and A.~Linde,
  Phys.\ Rev.\ D {\bf 91} (2015) 083528
  doi:10.1103/PhysRevD.91.083528
  [arXiv:1502.07733 [astro-ph.CO]].






\bibitem{linder} E.~V.~Linder,
  Phys.\ Rev.\ D {\bf 91} (2015) no.12,  123012
  doi:10.1103/PhysRevD.91.123012
  [arXiv:1505.00815 [astro-ph.CO]].

\bibitem{Odintsov:2016vzz}
  S.~D.~Odintsov and V.~K.~Oikonomou,
  Phys.\ Rev.\ D {\bf 94} (2016) no.12,  124026
  doi:10.1103/PhysRevD.94.124026
  [arXiv:1612.01126 [gr-qc]].



\bibitem{Odintsov:2016jwr}
  S.~D.~Odintsov and V.~K.~Oikonomou,
  Class.\ Quant.\ Grav.\  {\bf 34} (2017) no.10,  105009
  doi:10.1088/1361-6382/aa69a8
  [arXiv:1611.00738 [gr-qc]].




\bibitem{extra1} 
  R.~Kallosh, A.~Linde and D.~Roest,
  Phys.\ Rev.\ Lett.\  {\bf 112} (2014) no.1,  011303
  doi:10.1103/PhysRevLett.112.011303
  [arXiv:1310.3950 [hep-th]].

\bibitem{extra2} K.~Dimopoulos and C.~Owen,
  arXiv:1712.01760 [astro-ph.CO].

\bibitem{extra3} A.~Karam, T.~Pappas and K.~Tamvakis,
  Phys.\ Rev.\ D {\bf 96} (2017) no.6,  064036
  doi:10.1103/PhysRevD.96.064036
  [arXiv:1707.00984 [gr-qc]].

\bibitem{extra4} T.~Miranda, J.~C.~Fabris and O.~F.~Piattella,
  JCAP {\bf 1709} (2017) no.09,  041
  doi:10.1088/1475-7516/2017/09/041
  [arXiv:1707.06457 [gr-qc]].

\bibitem{extra5} G.~Narain,
  JCAP {\bf 1710} (2017) no.10,  032
  doi:10.1088/1475-7516/2017/10/032
  [arXiv:1708.00830 [gr-qc]].

\bibitem{extra6}  K.~Nozari and N.~Rashidi,
  Phys.\ Rev.\ D {\bf 95} (2017) no.12,  123518
  doi:10.1103/PhysRevD.95.123518
  [arXiv:1705.02617 [astro-ph.CO]].

\bibitem{extra7} K.~Dimopoulos and C.~Owen,
  JCAP {\bf 1706} (2017) no.06,  027
  doi:10.1088/1475-7516/2017/06/027
  [arXiv:1703.00305 [gr-qc]].


\bibitem{extra8}

 L.~Jarv, A.~Racioppi and T.~Tenkanen,
  arXiv:1712.08471 [gr-qc].


\bibitem{extra9}


T.~Markkanen, T.~Tenkanen, V.~Vaskonen and H.~Veermae,
  arXiv:1712.04874 [gr-qc].

\bibitem{extra10}

M.~Artymowski and J.~Rubio,
  Phys.\ Lett.\ B {\bf 761} (2016) 111
  doi:10.1016/j.physletb.2016.08.024
  [arXiv:1607.00398 [astro-ph.CO]].



\bibitem{extra11}

 G.~K.~Karananas and J.~Rubio,
  Phys.\ Lett.\ B {\bf 761} (2016) 223
  doi:10.1016/j.physletb.2016.08.037
  [arXiv:1606.08848 [hep-ph]].






\bibitem{starob1} A.~A.~Starobinsky,
  Phys.\ Lett.\ B {\bf 91} (1980) 99.
  doi:10.1016/0370-2693(80)90670-X



\bibitem{starob2}   J.~D.~Barrow and S.~Cotsakis,
  Phys.\ Lett.\ B {\bf 214} (1988) 515.
  doi:10.1016/0370-2693(88)90110-4







\bibitem{higgs} F.~L.~Bezrukov and M.~Shaposhnikov,
  Phys.\ Lett.\ B {\bf 659} (2008) 703
  doi:10.1016/j.physletb.2007.11.072
  [arXiv:0710.3755 [hep-th]].





\bibitem{Gao:2017uja}
  Q.~Gao and Y.~Gong,
  arXiv:1703.02220 [gr-qc].


\bibitem{Lin:2015fqa}
  J.~Lin, Q.~Gao and Y.~Gong,
  Mon.\ Not.\ Roy.\ Astron.\ Soc.\  {\bf 459} (2016) no.4,  4029
  doi:10.1093/mnras/stw915
  [arXiv:1508.07145 [gr-qc]].







\bibitem{Miranda:2017juz}
  T.~Miranda, J.~C.~Fabris and O.~F.~Piattella,
  JCAP {\bf 1709} (2017) no.09,  041
  doi:10.1088/1475-7516/2017/09/041
  [arXiv:1707.06457 [gr-qc]].


\bibitem{Fei:2017fub}
  Q.~Fei, Y.~Gong, J.~Lin and Z.~Yi,
  JCAP {\bf 1708} (2017) no.08,  018
  doi:10.1088/1475-7516/2017/08/018
  [arXiv:1705.02545 [gr-qc]].










\bibitem{MBojowald}
M. Bojowald,
Class.Quant.Grav. {\bf 26}, 075020 (2009)   [arXiv:0811.4129].


\bibitem{abl03}
 A.Ashtekar, M. Bojowald and J.Lewandowski, Adv.Theor.Math.Phys.
{\bf 7}, 233-
268 (2003) [arXiv:0304074].
\bibitem{bojowald05}
  M. Bojowald, Living Rev. Rel.
{\bf 8}, 11 (2005) [arXiv:0601085].


\bibitem{Singh07}
 P. Singh, J.Phys.Conf.Ser. {\bf 140}, 012005 (2008) [arXiv:0901.1301].



\bibitem{he}  J. Haro, E. Elizalde, EPL
{\bf 89},
69001 (2010).
\bibitem{dmp}  P. Dzierzak, P. Malkiewicz and W. Piechocki, Phys. Rev.
{\bf D 80}, 104001 (2009)
[arXiv:0907.3436].






\bibitem{martineau}
K. Martineau, A. Barrau and J. Grain,
{\it A first step towards the inflationary trans-planckian problem treatment in Loop Quantum Cosmology}
 [arXiv:1709.03301].

























\bibitem{Array:2015xqh}
  P.~A.~R.~Ade {\it et al.} [BICEP2 and Keck Array Collaborations],
  Phys.\ Rev.\ Lett.\  {\bf 116} (2016) 031302
  doi:10.1103/PhysRevLett.116.031302
  [arXiv:1510.09217 [astro-ph.CO]].















\bibitem{helling}
R. C. Helling,
{\it Higher curvature counter terms cause the bounce in loop cosmology}, (2009)
 [arXiv:0912.3011 [gr-qc]].



\bibitem{ds09}
G. Date and S. Sengupta,
\textit{Effective Actions from Loop Quantum Cosmology: Correspondence with Higher Curvature Gravity},
Class. Quant. Grav. {\bf 26}, 105002 (2009) [arXiv:0811.4023 [gr-qc]].

\bibitem{haro12}
 J. de Haro, \textit{Future singularity avoidance in phantom dark energy models}, JCAP
{\bf 1207}, 007 (2012) [arXiv:1204.5604 [gr-qc]].

\bibitem{w}
R. Weitzenb\"ock, {\it Invarianten Theorie},  Noordhoff, Groningen, (1923).





 \bibitem{ha17}
 J. de Haro and J. Amor\'os,  {\it Bouncing cosmologies via modified gravity in the ADM formalism: Application to Loop Quantum Cosmology},
 (2017) [arXiv:1712.08399 [gr-qc]].










\bibitem{cm}
 J.  Carminati and R. G. McLenaghan, \textit{Algebraic invariants of the Riemann tensor in a four-dimensional Lorentzian space}, J. Math. Phys. {\bf 32},  3135 (1991).


\bibitem{hp18}
J. de Haro and S. Pan, \textit{A note on bouncing backgrounds}, (2018)  [arXiv:1801.05475 [gr-qc]].


\bibitem{martineau1}
K. Martineau, A. Barrau and S. Schander,
    Phys. Rev. {\bf D 95}, 083507 (2017)    [arXiv:1701.02703].





\end{thebibliography}
\end{document}